\documentclass[journal,draftcls,onecolumn]{IEEEtran}

\IEEEoverridecommandlockouts

\usepackage[bibstyle=ieee,citestyle=numeric-comp]{biblatex}
\usepackage{mathrsfs,amsmath,amsfonts,amssymb}
\newtheorem{definition}{Definition}
\newtheorem{proposition}{Proposition}
\newtheorem{theorem}{Theorem}
\newtheorem{lemma}{Lemma}
\usepackage{graphicx}

\usepackage{algorithm}
\usepackage{algorithmic}

\usepackage{flushend}

\begin{document}
\title{Joint Offloading Decision and Resource Allocation for Vehicular Fog-Edge
Computing Networks: A Contract-Stackelberg Approach}

\author{Yuwei Li, Bo Yang, \IEEEmembership{Senior~Member,~IEEE,} Hao Wu,
Qiaoni Han, \\ Cailian Chen, \IEEEmembership{Member,~IEEE,} and Xinping Guan,
\IEEEmembership{Fellow,~IEEE}
\thanks{Copyright (c) 2022 IEEE. Personal use of this material is permitted.
However, permission to use this material for any other purposes must be obtained
from the IEEE by sending a request to pubs-permissions@ieee.org.}
\thanks{Y. Li was with the Department of Automation, Shanghai Jiao Tong
University, Shanghai 200240, China and is with Huawei Technologies Co., Ltd.
(email: liyuwei8@huawei.com).}
\thanks{B. Yang \emph{(Corresponding author)}, H. Wu, C. Chen, and X. Guan are
with the Department of Automation, Shanghai Jiao Tong University, Shanghai
200240, China; Key Laboratory of System Control and Information Processing,
Ministry of Education of China, Shanghai 200240, China; Shanghai Engineering
Research Center of Intelligent Control and Management, Shanghai 200240, China
(e-mail: bo.yang@sjtu.edu.cn; wuhao233@sjtu.edu.cn; cailianchen@sjtu.edu.cn;
xpguan@sjtu.edu.cn).  }
\thanks{Q. Han is with the Department of Automation, Tianjin University, Tianjin
300072, China (e-mail: qnhan@tju.edu.cn).}
\thanks{\\ DOI: 10.1109/JIOT.2022.3150955}
}

\maketitle

\begin{abstract}
    With the popularity of mobile devices and development of computationally
    intensive applications, researchers are focusing on offloading computation
    to Mobile Edge Computing (MEC) server due to its high computational
    efficiency and low communication delay.  As the computing resources of an
    MEC server are limited, vehicles in the urban area who have abundant idle
    resources should be fully utilized.  However, offloading computing tasks to
    vehicles faces many challenging issues.  In this paper, we introduce a
    vehicular fog-edge computing paradigm and formulate it as a multi-stage
    Stackelberg game to deal with these issues.  Specifically, vehicles are not
    obligated to share resources, let alone disclose their private information
    (\emph{e.g.}, stay time and the amount of resources).  Therefore, in the
    first stage, we design a contract-based incentive mechanism to motivate
    vehicles to contribute their idle resources.  Next, due to the complicated
    interactions among vehicles, road-side unit (RSU), MEC server and mobile
    device users, it is challenging to coordinate the resources of all parties
    and design a transaction mechanism to make all entities benefit.  In the
    second and third stages, based on Stackelberg game, we develop pricing
    strategies that maximize the utilities of all parties.  The analytical forms
    of optimal strategies for each stage are given.  Simulation results
    demonstrate the effectiveness of our proposed incentive mechanism, reveal
    the trends of energy consumption and offloading decisions of users with
    various parameters, and present the performance comparison between our
    framework and existing MEC offloading paradigm in vehicular networks.  
\end{abstract}
\begin{IEEEkeywords}
    Vehicular fog computing, mobile edge computing, computation offloading, on
    board unit, contract theory, Stackelberg game
\end{IEEEkeywords}

\section{Introduction}

\IEEEPARstart{T}{he} development of Internet of Things and wireless
communication technologies facilitates the emergence of computationally
intensive applications with requirements of low latency and real-time
processing, such as computer vision, natural language processing and autonomous
driving.  However, it is difficult for mobile devices with limited resources to
provide required quality of service (QoS) for users \cite{7486992}.  

Plenty of attempts have been made to offload computing tasks to the remote cloud
server \cite{7422842}.  Although cloud computing can significantly improve
computational performance, the long distance transmission may lead to
considerable overhead and large latency \cite{7271119}.  By contrast, MEC with
low cost servers in the vicinity of users, is a promising solution to
computation offloading.  

However, the computing resources of MEC server are usually constrained.
Offloading computationally intensive tasks to these servers may cause low QoS,
especially when the server is bursting with requests.  Therefore, it is urgent
to expand the resource capacity of MEC servers by exploiting idle resources from
existing network entities and then effectively scheduling them for better
services.  Besides, energy-efficient optimization is also required for
large-scale deployment and sustainable development of edge computing
\cite{7553459}.  

Recently, it is observed that vehicles with abundant idle computing resources
can be organized into a vehicular fog computing (VFC) network so as to improve
urban communication and computing capabilities \cite{7415983}.  In our daily
life, many vehicles stay in the parking lot for a long time, or move slowly on
the street.  Due to their large number, long-term stay or slow movement, and
relatively fixed location, vehicles in the urban area are ideal candidates for
static backbone network and edge servers.  In addition to the basic functions of
computation offloading, such as providing users with computing, storage and
application services, VFC also has the characteristics of low deployment cost,
proximity to users, dense geographical distribution and so on.  In this paper,
we propose to utilize urban vehicles' idle resources for computation offloading.

Nevertheless, there exist some challenging issues to be addressed.  Firstly,
vehicles are self-interested and have no obligation to share their idle
resources.  Therefore, a carefully designed incentive mechanism with reasonable
rewards is indispensable.  Additionally, vehicles are intuitively reluctant to
disclose their private information, which creates an information asymmetry
between vehicles and the organizer of VFC network like RSU.  Secondly, the
interactions among vehicles, RSU, the MEC server and mobile device users are
complicated and difficult to model, due to their rationality and selfishness,
and the tight coupling between resource demands and resource provisions.
Thirdly, in order to realize a real-time and scalable computation offloading
framework, an effective algorithm for joint offloading decision and resource
allocation is preferred.  

Motivated by the above issues, we introduce a new computing paradigm, named
vehicular fog-edge computing (VFEC).  In particular, a multi-stage Stackelberg
game with an incentive mechanism is proposed to model the VFEC scenario.  In a
long-term market, RSU rents computing resources from vehicles.  In this process,
the information superiority over RSU for vehicles hinders the realization of
Pareto optimal configuration.  Based on contract theory, we design incentive
mechanism to properly handle this problem.  By designing a series of contract
items for vehicles to choose from, the contract can reveal the respective types
of vehicles, make up for the influence of information asymmetry, and maximize
the utilities of two sides of trade.  On the other hand, the transaction between
RSU and the MEC server, as well as between the MEC server and users, can be
considered as a short-term market, which involves various interactions such as
publishing prices, offloading decision-making and computing resources purchase.
Stackelberg game is suitable for modelling this short-term, complex and dynamic
process.  Moreover, the solution to equilibrium of Stackelberg game only needs
one round of operations, instead of the iterative algorithms commonly used in
some auctions and non-cooperative games.  This is good news for the time-varying
computation offloading service market.  

The main contributions of this paper are listed below.  

\begin{IEEEitemize}
    \item We propose a vehicular fog-edge computing paradigm to utilize urban
        vehicles for computation offloading, and develop a multi-stage
        Stackelberg game with an incentive mechanism to model the interactions
        among RSU, MEC server and mobile device users.  
    \item We design a contract-based incentive mechanism for RSU to manage the
        idle computing resources of nearby vehicles.  In order to overcome the
        information asymmetry between vehicles and RSU, the contract items are
        designed carefully not only to maximize the utility of RSU, but also to
        satisfy individual rationality and incentive compatibility of vehicles.
    \item The analytical forms of optimal strategies for each stage of the
        transaction are given.  Numerical evaluation is conducted, which
        demonstrates the effectiveness of our proposed algorithms, reveal the
        trends of energy consumption and offloading decisions of users with
        various parameters.  
\end{IEEEitemize}

The rest of this paper is organized as follows.  Related works are listed in
Section \uppercase\expandafter{\romannumeral2}.  Section \uppercase\expandafter
{\romannumeral3} describes the system model.  Section \uppercase\expandafter
{\romannumeral4} introduces the multi-stage Stackelberg game with a
contract-based incentive mechanism.  The optimal algorithms are presented
elaborately in Section \uppercase\expandafter{\romannumeral5}.  The simulation
results are shown and interpreted in Section \uppercase\expandafter
{\romannumeral6}.  Finally, we conclude this paper in Section \uppercase
\expandafter{\romannumeral7}.  

\section{Related Works}

MEC has recently attracted widespread attention from academia and industry.  MEC
servers are located on the edge of wireless network with more computing and
storage resources than mobile terminals.  Many existing works focus on MEC
offloading in wireless networks, some of which are dedicated to minimize energy
consumption \cite{7553459,6574874,7130662,7762913,8496832}, reduce latency
\cite{7572018,7541539}, or optimize a weighted objective of energy consumption
and delay \cite{7870694,8533343}.  In addition, most of the above works optimize
resource allocation \cite{7762913} or joint computation offloading and resource
allocation \cite{7553459,6574874,7130662,7762913,8496832,7870694,8533343} for
MEC offloading in heterogeneous cellular networks.  Due to the roll out of 5G
mobile networks and Internet of Things, numerous and diverse users, servers and
applications coexist in MEC systems.  Therefore, server deployment and resource
allocation become quite complicated in such system.  Rodrigues \emph{et al.}
presented the utilization of Machine Learning (ML) to tackle these challenges in
\cite{8946743} and \cite{8847416}.  

Thanks to MEC's high computational efficiency and low communication delay, some
works have deployed MEC server in vehicular networks, and aim to better support
computationally intensive services with requirements of low latency and
real-time processing \cite{7981532}.  In \cite{liu2019vehicular}, Liu \emph{et
al.} provided an overview of Vehicular Edge Computing (VEC).  A mobility-aware
computation offloading design for MEC-based vehicular networks was studied in
\cite{8917559}.  Due to the limited computing resources of single MEC server and
high requirements for timely task processing of a large amount of computations
in the emerging mobile applications, some works focus on MEC cooperation or
grouping.  For instance, a cloud-MEC collaborative computation offloading scheme
in vehicular networks was presented in \cite{8745530}.  Considering the
cooperative utilization of computing resources of MEC/cloud servers, Dai
\emph{et al.} solved a distributed task assignment problem \cite{9091251}.
Based on game theory, a noncooperative game-based strategy selection algorithm
was presented in \cite{8936356} to realize MEC grouping for task offloading, and
a multi-user noncooperative computation offloading game was formulated in
\cite{8985335} to adjust the offloading probability of each vehicle.  Besides,
some works utilize neighbouring vehicles with idle computing resources to
provide offloading opportunities to other vehicles having limited computing
capabilities \cite{8753694,8436044,8581401}.  Among them, in \cite{8753694}, Gu
\emph{et al.} addressed the task offloading between MEC servers deployed at RSUs
and vehicles with excessive computing resources.  By treating vehicles as edge
computing infrastructure, Qiao \emph{et al.} introduced a vehicular edge
multi-access network and constructed a cooperative and distributed computing
architecture \cite{8436044}.  In \cite{8581401}, both an autonomous vehicular
edge (AVE) which shares neighbouring vehicles' available resources and a hybrid
vehicular edge cloud (HVC) which shares accessible resources of RSUs and cloud
were introduced.  Hou \emph{et al.} in \cite{9043503} proposed an
edge-computing-enabled software-defined Internet of vehicles (EC-SDIoV)
architecture to efficiently orchestrate the heterogeneous edge computing nodes,
and provide reliable and low-latency computation offloading, where partial
offloading, reliable task allocation and the reprocessing mechanism are jointly
considered.  In order to deal with the dynamic vehicular environment, Shi
\emph{et al.} in \cite{9277911} proposed a priority-aware task offloading scheme
in the context of vehicular fog computing, where a soft actor-critic method is
developed to select the service vehicles and publish dynamic prices.  The
algorithm in \cite{9277911} achieved more robust and sample-efficient
performance in task completion ratio and offloading delay.  More and deeper
issues of MEC are surveyed in \cite{8926369} toward future vehicular networks.  

Recently, parked and slow moving vehicles have attracted much attention to
improve the performance of vehicular networks.  Malandrino \emph{et al.}
investigated the possibility of exploiting parked vehicles to extend the RSU
service coverage \cite{6786409}.  In \cite{7747510}, a game theoretic framework
of content delivery which utilized parked vehicles was proposed.  Sun \emph{et
al.} considered the parked vehicles as relay nodes in \cite{8383688}.  Observing
that parked or slow moving vehicles have rich and under-utilized resources for
task execution, some works studied utilizing vehicles to develop new computing
paradigms \cite{7415983,grover2018vehicular}.  Aiming to minimize the average
response time for events reported by vehicles, Wang \emph{et al.} put forward a
feasible solution which enables offloading by moving and parked vehicles for
real-time traffic management \cite{8318667}.  With adopting parked vehicles for
computation offloading, an energy-efficient parked vehicular computing (PVC)
paradigm was developed in \cite{8463481}.  

Different from existing works, this paper focuses on some challenging issues.
First, vehicles have no obligation and willingness to share their idle computing
resources.  Second, there is information asymmetry between vehicles and the
organizer of VFC network, which makes it difficult to distinguish the
qualification of vehicles.  Therefore, it is necessary to design an incentive
mechanism to reveal the types of vehicles, and encourage them to participate in
the computation offloading with a certain reward.  Third, the offloading service
involves multiple transaction processes with multiple participants.  How to
coordinate the resources of all parties in an integrated way so that all
entities can benefit is also a major challenge.  In this paper, we introduce a
vehicular fog-edge computing paradigm and formulate it as a multi-stage
Stackelberg game to deal with these issues.  

\begin{table*}[!t]\tiny
    \renewcommand\arraystretch{1.5}
    \caption{Key Notations and Definitions}
    \label{tab:1}
    \centering
    \begin{tabular}{l l|l l}
        \hline
        Notation & Definition & Notation & Definition\\
        \hline
        $c_i$ & the computation needed by user $i$'s task & $f_i^{thresh}$ & a threshold of resources for judging whether user $i$ choose to offload its task or not \\
        $d_i$ & the size of input data of user $i$'s task & $f_e$ & the computing resources of MEC server used for task offloading \\
        $t_i^{\mathrm{max}}$ & the maximum tolerable latency of user $i$'s task & $f_{RSU}$ & the computing resources that the MEC server purchases from RSU \\
        $f_i^l$ & the local computing resources of user $i$ & $e, e_v$ & the unit monetary cost of server's / vehicle's energy consumption \\
        $f_e^{\mathrm{max}}$ & the computing resources of MEC server & $k$ & the effective switched capacitance, which depends on the chip architecture \\
        $f_v^{\mathrm{max}}$ & the idle computing resources of each vehicle & $\theta_m, m\in \{1, 2, \ldots, M\}$ & the type of a vehicle \\
        $U_i, U_{MEC}, U_{RSU}, U_{V_m}$ & the utility of user $i$ / the MEC server / RSU / a type-$m$ vehicle & $l_m$ & the number of type-$m$ vehicles near RSU \\
        $p$ & the price of MEC server's computing resources & $f_m$ & the computing resources that RSU rents from a type-$m$ vehicle \\
        $c$ & the price of computing resources collected by RSU & $p_m$ & the rent for leased resources of a type-$m$ vehicle per unit of time \\
        $f_i$ & the quantity of resources that user $i$ purchases from the MEC server & &\\
        \hline
    \end{tabular}
\end{table*}

\section{System Model}

\subsection{Overview}

We consider a small district in the urban area, where an MEC server, an RSU,
some vehicles and some mobile device users are located.  We assume that
the connections between all involved entities are single hop.  

There are $N$ mobile device users in this region, such as smart phones,
vehicles, wearable devices and so on.  We denote the ID set of these users as
$\mathscr{N} = \{1,2,\ldots,N\}$.  At some point, each user has a
computationally intensive task to be completed, which is characterized by a
tuple $(c_i, d_i, t_i^{\mathrm{max}}),\ \forall i \in \mathscr{N}$.  $c_i$
represents the computation (in CPU cycles) needed by user $i$'s task.  $d_i$ is
the size of input data of user $i$'s task.  $t_i^{\mathrm{max}}$ is the maximum
tolerable latency of user $i$.  $f_i^l$ denotes the local computing resources of
user $i$ (in CPU cycles/s).  

There exists an MEC server, whose computing resources, denoted as $f_e^
{\mathrm{max}}$, are limited.  A user may offload its task to the MEC server or
execute it locally, depending on the price of resources and how fast its task
will be completed.  

The resources of MEC server are not always able to meet the needs of all users,
which means high latency, expensive prices and low QoS.  We may consider making
full use of vehicular idle resources to improve users' experience, where the
resources are collected and scheduled by RSU.  

All vehicles in this region, whether parking or moving slowly, can establish
stable wireless connection with RSU via vehicle-to-infrastructure (V2I)
communications.  Each vehicle is equipped with an on-board-unit (OBU), which is
not a simple device tracking the vehicle location and measuring its speed, but a
mobile device with storage, communication and computing capabilities
\cite{9032323}.  We assume that the total amount of idle resources owned by each
vehicle, $f_v^{\mathrm{max}}$, is the same.  After signing the contract, OBUs
within the communication range of RSU can be organized into a fog cluster so as
to enhance service capabilities.  

In order to perform computation offloading efficiently after the MEC server has
prepared computing resources and users have made their offloading decisions, the
HVC architecture in \cite{8581401} may be used, which designs a workflow that
includes steps such as job caching, job scheduling, data transmission, job
execution and result data transmission.  Additionally, to facilitate the entire
transaction process and take care of the interests in all parties to a fair and
impartial manner, a third-party central controller is set up in the system,
which is subordinate to a computation offloading operator.  Its responsibilities
include collecting global information, running the proposed algorithms,
scheduling users' tasks and transmitting control messages, \emph{etc}.  As
mentioned earlier, this paper focuses on resource scheduling and users'
offloading decisions.  The details of these implementations are beyond the scope
of this paper and will be the direction of future work.  

\begin{figure}[!t]
    \centerline{\includegraphics[scale=0.25]{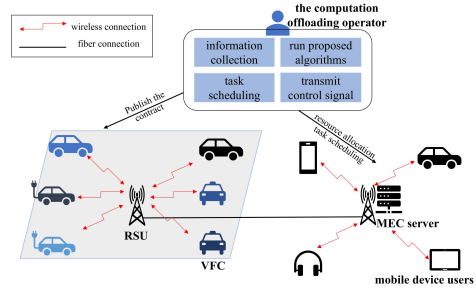}}
    \caption{The vehicular fog-edge computing network architecture}
\end{figure}

\subsection{Three-stage Stackelberg game}

In order to ensure that vehicles, RSU, the MEC server and users are willing to
participate in computation offloading, we design a multi-stage Stackelberg game
for this vehicular fog-edge computing paradigm.  In this game, a contract-based
incentive mechanism is developed for RSU to recruit the idle computing resources
of vehicles, while the interactions between RSU and the MEC server, the MEC
server and users are modeled as a multi-stage Stackelberg game.  

Specifically, in the first stage, a contract-based trading mechanism is designed
to employ vehicles for users' task execution.  The RSU acts as an employer who
offers different contract items to vehicles, while each vehicle is an employee
who selects a certain type of contract which suits it best.  At the same time,
RSU also acts as the leader of RSU-MEC Stackelberg game and announces the price
of computing resources to the MEC server.  

In the second stage, the MEC server is both the follower of RSU-MEC Stackelberg
game and the leader of MEC-user Stackelberg game, determining the quantity of
computing resources purchased from RSU and broadcasting the price of resources
to all mobile device users.  

In the third stage, all users are the followers of MEC-user Stackelberg game.
Each user decides whether to offload its task and the quantity of computing
resources purchased from the MEC server.  

Note that since parked or slow moving vehicles are considered, the leasing of
idle vehicular computing resources is a long-term market.  In contrast, the
deadline for a user's task is shorter, so the RSU-MEC Stackelberg game and the
MEC-user Stackelberg game are played slot by slot.  

\subsection{Utility functions}

\paragraph{mobile device users}
The utility function of user $i$ is defined as
\begin{equation}
    U_i(f_i) = \frac{C}{t_i^{\mathrm{max}}} \ln\left(\frac{f_i}{f^l_i} + \delta
    \right) - p f_i.
\end{equation}
In fact, users tend to purchase more computing resources to minimize delay.
However, with the continuous growth of obtained resources, the shortened
completion time is reduced, and the benefits of unit computing resource are
gradually declining as well.  That is, diminishing marginal effect appears.
Therefore, the benefits of a user can be characterized by a concave function of
the purchased computing resources over its local resources.  So we may adopt a
logarithmic function.  $C > 0$ is a constant, and $\tau_i \triangleq
C/t_i^{\mathrm{max}}$ represents user $i$'s sensitivity to latency.  $\delta \ge
1$, is to make sure that the logarithmic function has a positive value.  $p$
denotes the price of MEC server's computing resources, and $f_i$ is the quantity
of resources that user $i$ purchases from the MEC server.  The utility of a
mobile device user is defined as the benefits of offloading service minus the
expenditure of resources.  

If user $i$ decides to execute its task locally, the completion time is
\begin{equation}
    t_i^l = \frac{c_i}{f_i^l}.
\end{equation}
Otherwise, user $i$ offloads its task to the MEC server.  The completion time is
composed of upload latency and execution delay\footnote{Usually the results of
processing are much smaller than the input data, so here we ignore the download
transmission delay \cite{7572018}.}.  Then the completion latency for offloading
user $i$'s task to the MEC server is\footnote{In fact, a user's task may be
offloaded to a certain vehicle.  Then the upload latency in \eqref{eq:ct} should
be replaced by the upload delay from the user to the vehicle.  However, we can't
figure out in advance whether and which vehicle a task will be delivered to.
Additionally, considering that users, the MEC server and vehicles are all
located in the same small district, we use the upload delay from the user to the
MEC server to replace the specific transmission delay.}  
\begin{equation}\label{eq:ct}
    t_i^e = \frac{d_i}{r_i} + \frac{c_i}{f_i},
\end{equation}
where $r_i$ represents the upload transmission rate from user $i$ to the MEC
server.  When $t_i^e < t_i^l$, user $i$ may prefer to offload its
computationally intensive task to the MEC server.  Let $f_i^{thresh} =
\frac{c_i}{\frac{c_i}{f_i^l} - \frac{d_i}{r_i}}$.  If
\begin{equation}\label{eq:f_thresh}
    f_i > f_i^{thresh},
\end{equation}
then user $i$ will choose to offload its task.  

\paragraph{the MEC server}
The utility function of the MEC server is defined as
\begin{equation}
    U_{MEC}(p,f_e,f_{RSU}) = \sum_{i=1}^{N} p f_i - e k f_e^2 - c f_{RSU},
\end{equation}
where $f_e$ is the computing resources of MEC server used for task offloading,
and $e$ denotes the unit cost of server's energy consumption.  $k f_e^2$
represents the energy of MEC server consumed by computation.  Here, the energy
consumption model of computation is referenced from \cite{7572018}.  $k$ is the
effective switched capacitance, which depends on the chip architecture.  In
order to simplify expressions, let $k_e = e k$.  $c$ denotes the price of
resources collected by RSU, and $f_{RSU}$ is the computing resources that the
MEC server purchases from RSU.  The utility of MEC server is the revenues from
users minus the energy cost and the expenditure of purchasing computing
resources from the RSU.  

\paragraph{the RSU}
Before the contract is signed, we assume that RSU has obtained the current stay
time of each vehicle within its communication range, which is denoted as
$\theta_v^\prime$.  The type of a vehicle $v$ is defined as the length of time
it continues to park or stay, $\theta_v$.  Of course, RSU prefers the vehicle
who has a longer stay.  Due to the existence of information asymmetry between
vehicles and RSU, the latter does not know the exact types of former.  However,
through statistics on the historical data, RSU can obtain the probability
cumulative function of vehicle stay time $F(t)$ about the timing in a day,
vehicle position and other factors.  Then we have:
\begin{equation}
    P(\theta \ge \theta_v + \theta_v^\prime | \theta \ge \theta_v^\prime) =
    \frac{1-F(\theta_v + \theta_v^\prime)}{1 - F(\theta_v^\prime)}.
\end{equation}
In order to facilitate the subsequent analysis, we discretize the type of
vehicles into $M$ items: $\theta_1, \theta_2, \ldots, \theta_M$ with $\theta_1
\le \theta_2 \le \ldots \le \theta_M$.  Then it can be seen that RSU can
calculate the probability distribution of the vehicles' continued stay time at
the current moment, $f(\theta_m)$ with $\sum_{m=1}^M f(\theta_m) = 1$.
Therefore, the number of type-$m$ vehicles near the RSU is $l_m = f(\theta_m)L$,
where $L$ is the total number of vehicles.  

The utility function of RSU is
\begin{equation}
    U_{RSU}(f_m, p_m, c) = cf_{RSU} - \sum_{m=1}^M l_m \theta_m p_m,
\end{equation}
where $p_m$ represents the rent for leased resources of a type-$m$ vehicle per
unit of time.  The utility of RSU is defined as payment from the MEC server
minus the cost of renting computing resources from vehicles.  

\paragraph{vehicles}
The utility function of a type-$m$ vehicle is defined as
\begin{equation}
    U_{V_m}(f_m, p_m) = \theta_m p_m - e_v k f_m^2,
\end{equation}
where $f_m$ denotes the quantity of computing resources offered by a type-$m$
vehicle, and $e_v$ is the unit monetary cost of vehicle's energy consumption.
The utility of a vehicle is the rewards from RSU minus the cost of energy
consumption on computing.  

\section{Problem Formulation}

\subsection{Stage \uppercase\expandafter{\romannumeral1}}
To motivate the vehicles to participate the transaction and select the contract
item which fits their types best, the following IR and IC conditions should be
satisfied \cite{7997316}.  
\begin{definition}
    \textbf{[Individual Rationality (IR)]} Since the vehicles are rational, IR
    condition committed a nonnegative utility to a vehicle if it accepts the
    contract item designed for its type.  The IR conditions can be formulated as
    \begin{equation}\label{eq:IR}
        \theta_m p_m - k_v f_m^2 \ge 0,\ \forall m \in \{1,2,\cdots,M\}.
    \end{equation}
    Similar to the MEC server, let $k_v = e_v k$ to simplify the
    derivation process.  
\end{definition}
\begin{definition}
    \textbf{[Incentive Compatibility (IC)]} IC conditions guarantee that a
    type-$m$ vehicle will select the contract $(f_m, p_m)$, rather than any
    other contract items $(f_j, p_j),\ \forall j \in \{1,2,\cdots,M\}\backslash
    \{m\}$.  The IC conditions can be written as
    \begin{equation}\label{eq:IC}
        \theta_m p_m - k_v f_m^2 \ge \theta_m p_j - k_v f_j^2,\\
        \forall m,j \in \{1,2,\cdots,M\}, m\neq j.
    \end{equation}
\end{definition}

Meanwhile, the RSU also plays the leader of RSU-MEC Stackelberg game and
announces the price of computing resources $c$ to the MEC server.  We formulate
the utility maximization problem of RSU as
\begin{equation}\label{eq:P1}
    \begin{aligned}
        \max_{f_m, p_m, c}\quad &U_{RSU}(f_m, p_m, c) \\
        \mathrm{s.t.}\ \mathbf{C1:}\ &c \ge 0 \\
        \mathbf{C2:}\ &0\le f_m \le f_v^{\mathrm{max}},\ \forall m \in
        \{1,2,\cdots,M\} \\
        \mathbf{C3:}\ &\sum_{m=1}^M l_m f_m \ge f_{RSU} \\
        \mathbf{C4:}\ &\eqref{eq:IR} \\
        \mathbf{C5:}\ &\eqref{eq:IC}
    \end{aligned}
\end{equation}
\textbf{C3} ensures that the MEC server does not purchase more resources than
the resources RSU renting from vehicles.  

\subsection{Stage \uppercase\expandafter{\romannumeral2}}

In this transaction, the MEC server plays two roles.  It is the follower of
RSU-MEC Stackelberg game as well as the leader of MEC-user Stackelberg game.
The MEC server determines computing resources $f_{RSU}$ purchased from the RSU
and computing resources $f_e$ used locally based on the price of unit resources
$c$ of RSU.  At the same time, it broadcasts the price of unit resources $p$ to
all mobile device users.  We can describe the utility maximization problem of
MEC server as
\begin{equation}\label{eq:P2}
    \begin{aligned}
        \max_{p,f_e,f_{RSU}}\quad &U_{MEC}(p,f_e,f_{RSU}) \\
        \mathrm{s.t.}\ \mathbf{C6:}\ &p \ge 0 \\
        \mathbf{C7:}\ &0 \le f_e \le f_e^{\mathrm{max}} \\
        \mathbf{C8:}\ &f_{RSU} \ge 0 \\
        \mathbf{C9:}\ &f_e + f_{RSU} = \sum_{i=1}^N f_i
    \end{aligned}
\end{equation}

\subsection{Stage \uppercase\expandafter{\romannumeral3}}

The mobile device user $i$, as a follower of MEC-user Stackelberg game,
determines the quantity of computing resources $f_i$ purchased from the MEC
server based on the price $p$ to maximize its own utility.  We can formulate the
optimization problem of user $i$ as
\begin{equation}\label{eq:P3}
    \begin{aligned}
        \max_{f_i}\quad &U_i(f_i) \\
        \mathrm{s.t.}\ \mathbf{C10:}\ &\eqref{eq:f_thresh}
    \end{aligned}
\end{equation}

\section{Optimal Algorithm}

In this section, we use the backward induction method to analyze this
multi-stage Stackelberg game and the contract-based incentive mechanism.  

\subsection{Solution of Stage \uppercase\expandafter{\romannumeral3}}

Since the utility function of user $i$ is a concave function, the zero point of
its first derivative is the optimal solution.  That is,
\begin{equation}\label{eq:optimal-f_i}
    f_i^\star = \begin{cases}
        \frac{\tau_i}{p} - f_i^l \delta,\quad &p < \frac{\tau_i}{f_i^l\delta +
        f_i^{thresh}} \\
        0,\ &p \ge \frac{\tau_i}{f_i^l\delta + f_i^{thresh}}
    \end{cases}
\end{equation}
We can see that if the price of computing resources is too high, \emph{i.e.},
$p \ge \frac{\tau_i}{f_i^l\delta + f_i^{thresh}}$, user $i$ is unwilling to
offload its task.  

\subsection{Solution of Stage \uppercase\expandafter{\romannumeral2}}

Given the quantity of computing resources \eqref{eq:optimal-f_i} purchased by
users, we can substitute it into the utility function of MEC server.  Note that
$f_i^\star$ is a piecewise function, we introduce the following indicator
variable for user $i$
\begin{equation}\label{eq:indicator-chi}
    \chi_i = \begin{cases}
        1,\quad &p < \frac{\tau_i}{f_i^l\delta + f_i^{thresh}} \\
        0,\quad &p \ge \frac{\tau_i}{f_i^l\delta + f_i^{thresh}}
    \end{cases}
\end{equation}
Then \eqref{eq:P2} can be rewritten as
\begin{equation}\label{eq:P4}
    \begin{aligned}
        \max_{p,f_e,f_{RSU},\chi}\quad &\sum_{i=1}^N \chi_i \left(\tau_i - f_i^l
        \delta p\right) - k_e f_e^2 - c f_{RSU} \\
        \mathrm{s.t.}\ \mathbf{C6} &\mathbf{-C9} \\
        \mathbf{C11}&\mathbf{:}\ \chi_i \in \{0,1\},\ \forall i \in \mathscr{N}
    \end{aligned}
\end{equation}
where $\chi = [\chi_1,\chi_2,\cdots,\chi_N]^\intercal$.  Since $\chi$ is a
binary vector, and $p, f_e, f_{RSU}$ are continuous variables, the optimization
problem \eqref{eq:P4} is a mixed integer nonlinear programming problem.  Given
the indicator vector $\chi$, it is easy to verify that this problem is concave.

We assume that the users are sorted in the following order:
\begin{equation}\label{eq:sort-users}
    \begin{aligned}
        \frac{\tau_1}{f_1^l \delta + f_1^{thresh}} & \ge \frac{\tau_2}{f_2^l
        \delta + f_2^{thresh}} \\
        & \ge \cdots \\
        & \ge \frac{\tau_N}{f_N^l \delta + f_N^{thresh}}.
    \end{aligned}
\end{equation}
According to \eqref{eq:indicator-chi}, we can see that whether a user chooses to
offload its task depends on the price of unit computing resources of MEC server,
\emph{i.e.}, $p$.  Further, in our transaction framework, $p$ depends on the
price of unit computing resources of RSU, \emph{i.e.}, $c$.  Thus, we consider a
special case of \eqref{eq:P4}, in which we assume that $c$ is small enough such
that all users choose to participate in the offloading.  That is, the indicators
for all users are equal to $1$.  Under this assumption, \eqref{eq:P4} can be
rewritten as
\begin{equation}\label{eq:P5}
    \begin{aligned}
        \max_{p,f_e,f_{RSU}}\quad &\sum_{i=1}^N \left(\tau_i-f_i^l\delta p
        \right) - k_e f_e^2 - c f_{RSU} \\
        \mathrm{s.t.}\ \mathbf{C6} &\mathbf{-C8}\ \\
        \mathbf{\overline{C9}}&\mathbf{:}\ \sum_{i=1}^N \left(\frac{\tau_i}{p} -
        f_i^l \delta\right) = f_e + f_{RSU}
    \end{aligned}
\end{equation}

\begin{proposition}
    The optimal solution of \eqref{eq:P5} is
    \begin{equation}\label{eq:optimal-p}
        p^\star = \sqrt{\frac{c \sum_{i=1}^{N} \tau_i} {\delta \sum_{i=1}^{N}
        f_i^l}},
    \end{equation}
    \begin{equation}
        f_e^\star = \min\left\{\frac{c}{2k_e}, f_e^{\mathrm{max}}\right\},
    \end{equation}
    when $2 k_e f_e^{\mathrm{max}} \le T_N$, $f_{RSU}^\star$ is shown as
    \begin{equation}\label{eq:optimal-f_RSU-1}
        f_{RSU}^\star = \begin{cases}
            \left[\sqrt{\frac{\tau\Delta}{c}} -\Delta - \frac{c}{2k_e}\right]^+
            ,\ &c \le 2 k_e f_e^{\mathrm{max}} \\
            \left[\sqrt{\frac{\tau\Delta}{c}} - \Delta - f_e^{\mathrm{max}}
            \right]^+ ,\ &2 k_e f_e^{\mathrm{max}} < c < T_N
        \end{cases}
    \end{equation}
    when $2 k_e f_e^{\mathrm{max}} > T_N$,
    \begin{equation}\label{eq:optimal-f_RSU-2}
        f_{RSU}^\star = \left[\sqrt{\frac{\tau\Delta}{c}} - \Delta - \frac{c}{2
        k_e}\right]^+,\ 0 \le c < T_N,
    \end{equation}
    where $\tau = \sum_{i=1}^N \tau_i, \Delta = \sum_{i=1}^N f_i^l \delta, T_N =
    \frac{\tau_N\sqrt{\delta \sum_{i=1}^N f_i^l}}{\left(f_N^l \delta +
    f_N^{thresh}\right) \left(\sum_{i=1}^N \tau_i\right)}$.  The proof of this
    proposition is shown in part A of Appendix.  
\end{proposition}
\begin{proposition}
    The price of MEC server's computing resources given by \eqref{eq:optimal-p}
    and the quantity of computing resources purchased by the MEC server from RSU
    given by \eqref{eq:optimal-f_RSU-1} and \eqref{eq:optimal-f_RSU-2} are
    optimal solution of \eqref{eq:P4} \emph{if and only if} $c < T_N$.  The
    proof of Proposition 2 is shown in part B of Appendix.  
\end{proposition}

Then the optimal solution of \eqref{eq:P4} is given by the following theorem.
\begin{theorem}
    The optimal solution of \eqref{eq:P4} is
    \begin{equation}\label{eq:optimal-p2}
        p^\star = \begin{cases}
            q_N,\quad &c < T_N \\
            q_{N-1},\quad &T_N \le c < T_{N-1} \\
            \quad\vdots\quad &\quad\vdots \\
            q_1,\quad & T_2 \le c < T_1
        \end{cases}
    \end{equation}

    (\romannumeral1): $2k_e f_e^{\mathrm{max}} \ge T_1$,
    \begin{equation}\label{eq:optimal-f_RSU1}
        f_{RSU}^\star = \begin{cases}
            \left[L_N - \frac{c}{2 k_e}\right]^+,\quad &c < T_N \\
            \left[L_{N-1}-\frac{c}{2 k_e}\right]^+,\quad &T_N \le c < T_{N-1} \\
            \quad\vdots\quad &\quad\vdots \\
            \left[L_1 - \frac{c}{2k_e}\right]^+,\quad &T_2 \le c < T_1
        \end{cases}
    \end{equation}

    (\romannumeral2): $2k_e f_e^{\mathrm{max}} \le T_N$
    \begin{equation}
        f_{RSU}^\star = \begin{cases}
            \left[L_N - \frac{c}{2 k_e}\right]^+,\quad &c\le 2 k_e
            f_e^{\mathrm{max}} \\
            \left[L_N - f_e^{\mathrm{max}}\right]^+,\quad &2k_e
            f_e^{\mathrm{max}} < c < T_N \\
            \left[L_{N-1} - f_e^{\mathrm{max}}\right]^+,\quad &T_N \le c <
            T_{N-1} \\
            \quad \vdots\quad &\quad\vdots \\
            \left[L_1 - f_e^{\mathrm{max}}\right]^+,\quad &T_2 \le c < T_1
        \end{cases}
    \end{equation}

    (\romannumeral3): $T_N < 2 k_e f_e^{\mathrm{max}} < T_1$ and $T_o \le 2k_e
    f_e^{\mathrm{max}} < T_{o-1}, \forall o \in \{2,3,\cdots,N\}$
    \begin{equation}
        f_{RSU}^\star = \begin{cases}
            \left[L_N - \frac{c}{2k_e}\right]^+,\quad &c < T_N \\
            \quad\vdots\quad &\quad\vdots \\
            \left[L_o - \frac{c}{2k_e}\right]^+,\quad &T_o \le c \le 2 k_e
            f_e^{\mathrm{max}} \\
            \left[L_o - f_e^{\mathrm{max}}\right]^+,\quad &2k_e
            f_e^{\mathrm{max}} < c < T_{o-1} \\
            \quad\vdots\quad &\quad\vdots \\
            \left[L_1 - f_e^{\mathrm{max}}\right]^+,\quad &T_2 \le c < T_1
        \end{cases}
    \end{equation}
    where $q_k = \sqrt{\frac{c \sum_{i=1}^k \tau_i}{\delta \sum_{i=1}^k f_i^l}},
    T_k = \frac{\tau_k \sqrt{\delta \sum_{i=1}^k f_i^l}} {\left(f_k^l\delta +
    f_k^{thresh}\right) \left(\sum_{i=1}^k \tau_i \right)}, L_k =
    \sqrt{\frac{\left(\sum_{i=1}^k \tau_i \right)\left(\delta \sum_{i=1}^k
    f_i^l\right)}{c}} - \delta \sum_{i=1}^k f_i^l,\ \forall k \in
    \{1,2,\cdots,N\}$.
\end{theorem}

\begin{IEEEproof}
    If $c < T_N$, the optimal solution of \eqref{eq:P4} is obtained by
    Proposition 2.  For $c$ in other intervals, the solution can be obtained
    similarly as Proposition 2, and is thus omitted.  
\end{IEEEproof}
The algorithm for optimal solution of Stage \uppercase\expandafter
{\romannumeral2} is shown as below:

\begin{algorithm}
    \caption{the optimal solution of Stage \uppercase\expandafter
    {\romannumeral2}}
    \begin{algorithmic}
        \STATE \textit{Step 1}: Initialize $k = N$;
        \STATE \textit{Step 2}: Sort all users according to \eqref{eq:sort-users};
        \STATE \textit{Step 3}: Compute $T_k = \frac{\tau_k \sqrt{\delta
        \sum_{i=1}^k f_i^l}} {\left(f_k^l\delta + f_k^{thresh}\right)
        \left(\sum_{i=1}^k \tau_i \right)}$;
        \STATE \textit{Step 4}: Compare $T_k$ with $c$.  If $c \ge T_k$, remove
        user $k$ from the game, set $k = k - 1$, and go to step 3; otherwise, go
        to step 5;
        \STATE \textit{Step 5}: According to Theorem 1, find the optimal
        solution $p^\star$ and $f_{RSU}^\star$.
    \end{algorithmic}
\end{algorithm}

\subsection{Solution of Stage \uppercase\expandafter{\romannumeral1}}

It can be seen that the expression of $f_{RSU}^\star$ can be divided into three
cases depending on the relationship between the values of $2k_e
f_e^{\mathrm{max}}$ and $T_k, \forall k \in \{ 1,2,\cdots,N\}$.  Here, we only
give the solving process of Stage \uppercase\expandafter{\romannumeral1} in the
case of $2k_e f_e^{\mathrm{max}} \ge T_1$.  The other two cases can be solved in
similar ways.  To solve the utility maximization problem of RSU \eqref{eq:P1},
we substitute \eqref{eq:optimal-f_RSU1} into \eqref{eq:P1}.  Since the
expression of $f_{RSU}^\star$ is piecewise, the problem is decomposed into $N$
subproblems.  Assuming that $T_k \le c < T_{k-1}$, \eqref{eq:P1} can be
rewritten as:
\begin{equation}\label{eq:P6}
    \begin{aligned}
        \min_{f_m,p_m,c}\quad &\sum_{m=1}^M l_m \theta_m p_m - c \left(L_k -
        \frac{c}{2 k_e}\right) \\
        \mathrm{s.t.}\ \mathbf{C2,}&\mathbf{C4,C5} \\
        \mathbf{\overline{C1}:}\ &T_k \le c < T_{k-1} \\
        \mathbf{\overline{C3}:}\ &L_k-\frac{c}{2 k_e} \le \sum_{m=1}^M l_m f_m \\
        \mathbf{C11:}\ &L_k - \frac{c}{2 k_e} \ge 0
    \end{aligned}
\end{equation}

The optimization problem \eqref{eq:P6} is difficult to solve due to $M$ IR
constraints and $M(M-1)$ IC constraints.  Below we propose several lemmas to
simplify these constraints.  

\begin{lemma}
    For any feasible contract, if $\theta_i > \theta_j$, then we have $p_i >
    p_j, \forall i,j \in \{1,2,\cdots,M\}$.  
\end{lemma}

\begin{lemma}
    For any feasible contract, $p_i > p_j$ if and only if $f_i > f_j, \forall
    i,j \in \{1,2,\cdots,M\}$.  
\end{lemma}

The above two lemmas have been proved in \cite{7997316}.  Then, we can reduce IR
and IC constraints by the following lemmas.  

\begin{lemma}
    If the IR constraint of type-$1$ vehicles is satisfied, then all other IR
    constraints of type-$m$, $m \in \{2,3,\cdots,M\}$ vehicles are also
    satisfied.  That is,
    \begin{equation}
        \theta_m p_m - k_v f_m^2 \ge \theta_1 p_1 - k_v f_1^2 \ge 0.
    \end{equation}
\end{lemma}

\begin{lemma}
    The IC conditions can be reduced to the local downward incentive
    compatibility (LDIC) conditions and the local upward incentive compatibility
    (LUIC):
    \begin{equation}
        \theta_m p_m - k_v f_m^2 \ge \theta_m p_{m-1} - k_v f_{m-1}^2,\ \forall
        m \in \{2,3,\cdots,M\},
    \end{equation}
    \begin{equation}
        \theta_m p_m - k_v f_m^2 \ge \theta_m p_{m+1} - k_v f_{m+1}^2,\ \forall
        m \in \{1,2,\cdots,M-1\}.
    \end{equation}
\end{lemma}

\begin{lemma}
    Under the optimal contract, all the LDIC constraints are active, and the IR
    constraint of type-$1$ vehicles is active as well.  That is,
    \begin{equation}
        \theta_m p_m - k_v f_m^2 = \theta_m p_{m-1} - k_v f_{m-1}^2,\ \forall m
        \in \{2,3,\cdots,M\},
    \end{equation}
    \begin{equation}
        \theta_1 p_1 - k_v f_1^2 = 0.
    \end{equation}
\end{lemma}

\begin{lemma}
    If all the LDIC constraints are active, then all the LUIC constraints are
    satisfied.  
\end{lemma}

See the Appendix for proofs of above four lemmas.  

Then, the IR and IC conditions are reduced to active type-$1$ IR constraint and
active LDIC constraints.  Note that $\mathbf{C11}$ is not concave, we
transform $\mathbf{C11}$ to a concave function of $c$ $\mathbf{\overline{C11}}$.
Thus, \eqref{eq:P6} can be rewritten as
\begin{equation}\label{eq:P7}
    \begin{aligned}
        \min_{f_m,p_m,c}\quad &\sum_{m=1}^M l_m\theta_m p_m - c\left(L_k -
        \frac{c}{2 k_e}\right) \\
        \mathrm{s.t.}\ \mathbf{\overline{C1},}&\mathbf{C2,\overline{C3}} \\
        \mathbf{\overline{C4}:}\ &\theta_1 p_1 - k_v f_1^2 = 0 \\
        \mathbf{\overline{C5}:}\ &\theta_m p_m - k_v f_m^2 = \theta_m p_{m-1} -
        k_v f_{m-1}^2,\ m \ge 2 \\
        \mathbf{\overline{C11}:}\ &c\left(L_k - \frac{c}{2 k_e}\right) \ge 0 \\
        \mathbf{C12:}\ &0 \le p_1 \le p_2 \le \cdots \le p_M
    \end{aligned}
\end{equation}

Obviously, the optimization problem \eqref{eq:P7} is convex and can be solved by
Lagrangian multiplier method.  The Lagrangian function is defined as
\begin{equation}\label{eq:Lagrangian}
    \begin{aligned}
        \mathscr{L} = &\sum_{m=1}^M l_m\theta_m p_m - c\left(L_k - \frac{c}{2
        k_e}\right) \\
        +&\gamma(\theta_1 p_1 - k_v f_1^2) + \beta \left(-c L_k + \frac{c^2}{2
        k_e}\right) \\
        +&\sum_{m=2}^M \mu_m \left(\theta_m(p_{m-1} - p_m) - k_v(f_{m-1}^2 -
        f_m^2)\right) \\
        +&\eta\left(L_k - \frac{c}{2 k_e} - \sum_{m=1}^M l_m f_m\right),
    \end{aligned}
\end{equation}
where $\gamma, \beta, \mu = \{\mu_m\}, \eta$ are Lagrangian multipliers for
constraints $\mathbf{\overline{C4},\ \overline{C11},\ \overline{C5}}$ and
$\mathbf{\overline{C3}}$.  $\mathbf{C12}$ is used to verify feasibility of the
obtained optimal $p_m, \forall m \in \{1,2,\cdots,M\}$.  Then, we can get the
optimal solution presented in Theorem 2.  

\begin{theorem}
    The optimal solution of \eqref{eq:P7} is given as
    \begin{equation}\label{eq:mu_m}
        \mu_m = \frac{\mu_{m+1}\theta_{m+1}}{\theta_m} + l_m, \forall m \in \{
        2,3,\cdots,M-1\}, \mu_M = l_M,
    \end{equation}
    \begin{equation}\label{eq:f_1}
        f_1 = \frac{l_1 \eta}{2 k_v \left(\l_1 + \mu_2 \frac{\theta_2 -
        \theta_1}{\theta_1}\right)},
    \end{equation}
    \begin{equation}\label{eq:f_m}
        f_m = \frac{l_m \eta}{2 k_v (\mu_m - \mu_{m+1})},\ \forall m \in \{
        2,3,\cdots,M-1\},
    \end{equation}
    \begin{equation}\label{eq:f_M}
        f_M = \frac{\eta}{2 k_v},
    \end{equation}
    \begin{equation}\label{eq:p_1}
        p_1 = \frac{k_v f_1^2}{\theta_1},
    \end{equation}
    \begin{equation}\label{eq:p_m}
        p_m = p_{m-1} + \frac{k_v \left(f_m^2 - f_{m-1}^2\right)}{\theta_m},\
        \forall m \in \{2,3,\cdots,M\}.
    \end{equation}
\end{theorem}
The proof of Theorem 2 is shown in part G of Appendix.  

The subgradient method is used to update Lagrangian multipliers $\eta$ and
$\beta$.
\begin{equation}\label{eq:update_multipliers}
    \begin{aligned}
        \eta^{(l+1)} &= \left[\eta^{(l)} + k^{(l)}\left(L_k^{(l)} -\frac{c^{(l)}}
        {2 k_e} - \sum_{m=1}^M l_m f_m^{(l)}\right)\right]^+, \\
        \beta^{(l+1)} &= \left[\beta^{(l)} + k^{(l)}\left(\frac{{c^{(l)}}^2}{2
        k_e} - c^{(l)} L_k^{(l)}\right)\right]^+,
    \end{aligned}
\end{equation}
where $k^{(l)} = \frac{q}{\sqrt{l}}$ is the learning rate, $l$ is the number of
iterations, and $q > 0$ is a small positive constant.  

The optimal solution $f_m^\star, p_m^\star$ and $c^\star$ of Stage
\uppercase\expandafter{\romannumeral1} is selected from the optimal solutions of
$N$ optimization subproblems, \emph{i.e.}, the one which makes $U_{RSU}$
largest.  The other two cases $T_N < 2 k_e f_e^{\mathrm{max}} < T_1$ and $2 k_e
f_e^{\mathrm{max}} \le T_N$ can be solved in similar ways.  Lagrangian
multiplier method for Stage \uppercase\expandafter{\romannumeral1} is shown in
Algorithm 2.  

\begin{algorithm}
    \caption{Lagrangian multiplier method for the Stage \uppercase\expandafter
    {\romannumeral1}}
    \begin{algorithmic}[1]
        \STATE Initialize Lagrangian multipliers $\eta^{(l)}, \beta^{(l)}$, set
        $l = 0, \mu_M = l_M$;
        \FOR{$m = M - 1$ to $2$}
        \STATE $\mu_m = \frac{\mu_{m+1}\theta_{m+1}}{\theta_m} + l_m$
        \ENDFOR

        \REPEAT
        \STATE $f_1^{(l)} = \frac{l_1\eta^{(l)}}{2 k_v \left(l_1 + \mu_2
        \frac{\theta_2 - \theta_1}{\theta_1}\right)}$
        \STATE $f_M^{(l)} = \frac{\eta^{(l)}}{2 k_v}$
            \FOR{$m = M-1$ to $2$}
            \STATE $f_m^{(l)} = \frac{l_m \eta^{(l)}}{2 k_v (\mu_m -
            \mu_{m+1})}$
            \ENDFOR

            \STATE $p_1^{(l)} = \frac{k_v f_1^{(l)2}}{\theta_1}$

            \FOR{$m = 2$ to $M$}
            \STATE $p_m^{(l)} = p_{m-1}^{(l)} + \frac{k_v (f_m^{(l)2} -
            f_{m-1}^{(l)2})} {\theta_m}$
            \ENDFOR
            \STATE $c^{(l)} = \left[c^{(l-1)} - \frac{\partial L}{\partial
            c^{(l-1)}}\right]_{T_k}^{T_{k-1}}$
            \STATE Update $\eta^{(l+1)}, \beta^{(l+1)}$ according to
            \eqref{eq:update_multipliers};
            \STATE $l = l + 1$
        \UNTIL{$U_{RSU}$ converges}
    \end{algorithmic}
\end{algorithm}

As already proved in \cite{1665007}, the convergence of Algorithm 2 can be
guaranteed by adopting decreasing step sizes $k^{(l)}$ for
\eqref{eq:update_multipliers}.  Since we define $k^{(l)} = \frac{q}{\sqrt{l}}$,
the conditions in \cite{1665007}, \emph{i.e.}, $\sum_{l=1}^\infty k^{(l)} =
\infty$ and $\sum_{l=1}^\infty (k^{(l)})^2 < \infty$, are satisfied.  Therefore,
Algorithm 2 converges to the optimal solution given in Theorem 2.  

The algorithm for optimal solution of the multi-stage Stackelberg game and the
contract-based incentive mechanism is shown below.

\subsection{The computational complexity analysis}

\paragraph{Stage \uppercase\expandafter{\romannumeral1}}

The comparison of $T_k$ and $2 k_e f_e^{\mathrm{max}}$ will take $O(N)$ time.
According to \textit{Theorem 2} and \eqref{eq:update_multipliers}, we can see
that the time complexity of one-step iteration of primal variables and
Lagrangian multipliers is $O(M)$.  We assume that when $\frac{\Vert
U_{RSU}^{(l+1)} - U_{RSU}^{(l)}\Vert}{\Vert U_{RSU}^{(l)} - U_{RSU}^{(l-1)}
\Vert} \le \varepsilon$, Lagrangian multiplier method will stop.  Then it can be
seen that the number of iterations $l$ is $O(\mathrm{ln} \varepsilon)$
\cite{bertsekas1997nonlinear}.  Since the optimal solution of Stage \uppercase
\expandafter{\romannumeral1} is selected from the optimal solutions of $N$
optimization subproblems, the total time complexity of this Stage is $O(N + NM
\mathrm{ln}\varepsilon) = O(NM\mathrm{ln}\varepsilon)$.  

\paragraph{Stage \uppercase\expandafter{\romannumeral2}}

In the worst case, the time complexity of Algorithm 1 depends on \textit{Step 3}
and \textit{Step 4}, namely $O(N^2)$.  

\paragraph{Stage \uppercase\expandafter{\romannumeral3}}

According to \eqref{eq:optimal-f_i}, the time complexity of solving $f_i^\star$
is a constant, $O(1)$.  

Based on the above analysis, our proposed algorithm is a polynomial-time
algorithm, which is efficient.  

\begin{algorithm}
    \caption{the algorithm for optimal solution of the multi-stage Stackelberg
    game and the contract-based incentive mechanism}
    \begin{algorithmic}
        \STATE \textit{Step 1}: Compare $T_k, \forall k \in \{1,2,\cdots,N\}$
        with $2 k_e f_e^{\mathrm{max}}$ and use Algorithm 2 or similar methods
        to obtain the RSU's optimal solution $f_m^\star, p_m^\star$ and
        $c^\star, \forall m \in \{1,2,\cdots,M\}$;
        \STATE \textit{Step 2}: Use Algorithm 1 to find the MEC server's optimal
        solution $f_{RSU}^\star$ and $p^\star$;
        \STATE \textit{Step 3}: According to \eqref{eq:indicator-chi}, user $i$
        determines whether to offload its task to the MEC server.  If yes, the
        computing resources purchased by user $i$ is obtained based on
        \eqref{eq:optimal-f_i}.
    \end{algorithmic}
\end{algorithm}

\section{Numerical Results}

\subsection{Parameter Setting}

In this section, we use numerical simulation to evaluate the performance of our
proposed algorithm.  In the simulation, we set $\delta = 1, C = 100$.  We assume
that there are $M = 10$ types of vehicles in this $200 \mathrm{m} \times
200\mathrm{m}$ district. The probability of vehicles' types and some other
parameters are borrowed from \cite{8463481}.  We set the time interval $t = 20\
\mathrm{min}$ and the value of type $m$ is defined as $\theta_m = (m \cdot t) /
60, \forall m \in \{1,2,\cdots,M\}$.  The total computing resources of MEC
server is $f_e^{\mathrm{max}} = 300\mathrm{GHz}$, while each user and vehicle
has $1 \mathrm{GHz}$ local computing resources, \emph{i.e.}, $f_i^l =
f_v^{\mathrm{max}} = 1 \mathrm{GHz}, \forall i \in \mathscr{N}$.  The maximum
delay tolerance of user $i$, $t_i^{\mathrm{max}}$ is assigned as a random number
in $[0.5,\ 2]\ \mathrm{ms}$.  

\subsection{Convergence and Effectiveness}

\begin{figure}[!t]
    \centerline{\includegraphics[scale=0.6]{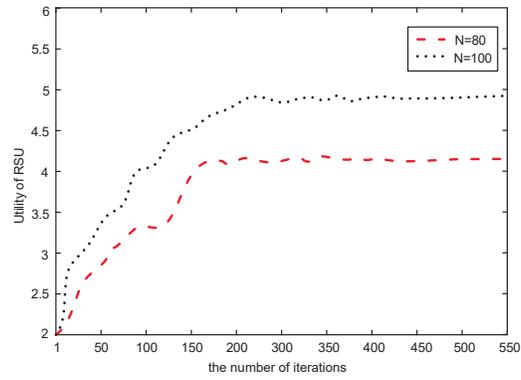}}
    \caption{The convergence of utility of RSU with different number of
    users}
    \label{fig:3}
\end{figure}

\begin{figure}[!t]
    \centerline{\includegraphics[scale=0.55]{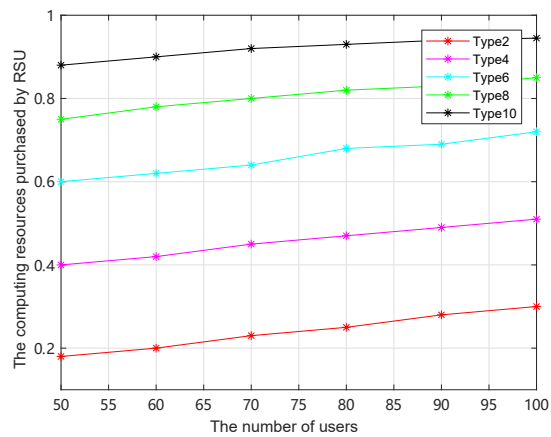}}
    \caption{The computing resources purchased by RSU with different number of
    users}
    \label{fig:4}
\end{figure}

\begin{figure}[!t]
    \centerline{\includegraphics[scale=0.55]{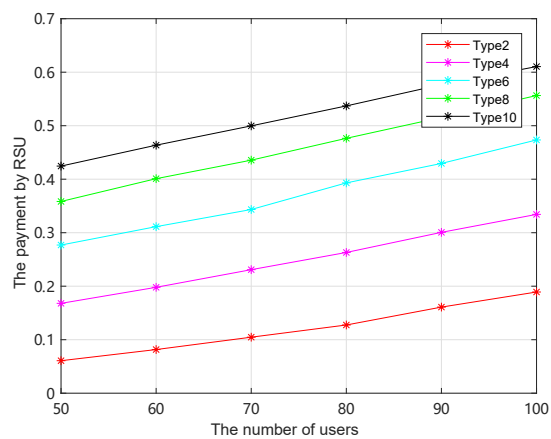}}
    \caption{The payment by RSU with different number of users}
    \label{fig:5}
\end{figure}

With $e = e_v = a = 60$, Fig. \ref{fig:3} shows the convergence of utility of
RSU with different number of users, and demonstrates the convergence of
Algorithm 2.  Fig. \ref{fig:4} and Fig. \ref{fig:5} show the optimal contracts
between the RSU and vehicles.  As shown in Fig. \ref{fig:4}, with the number of
mobile device users rising, the computing resources that the RSU borrows from
different types of vehicles also rise.  Fig. \ref{fig:4} verifies Lemma 2,
showing that the higher type a vehicle is, the more computing resources the RSU
purchases.  Fig. \ref{fig:5} shows that when the computing resources purchased
by RSU increase, the expenses paid by RSU to each type of vehicles also
increase.  Fig. \ref{fig:5} demonstrates Lemma 1 as well, showing that the
higher type a vehicle is, the more payment it gets.  

\begin{figure}[!t]
    \centerline{\includegraphics[scale=0.55]{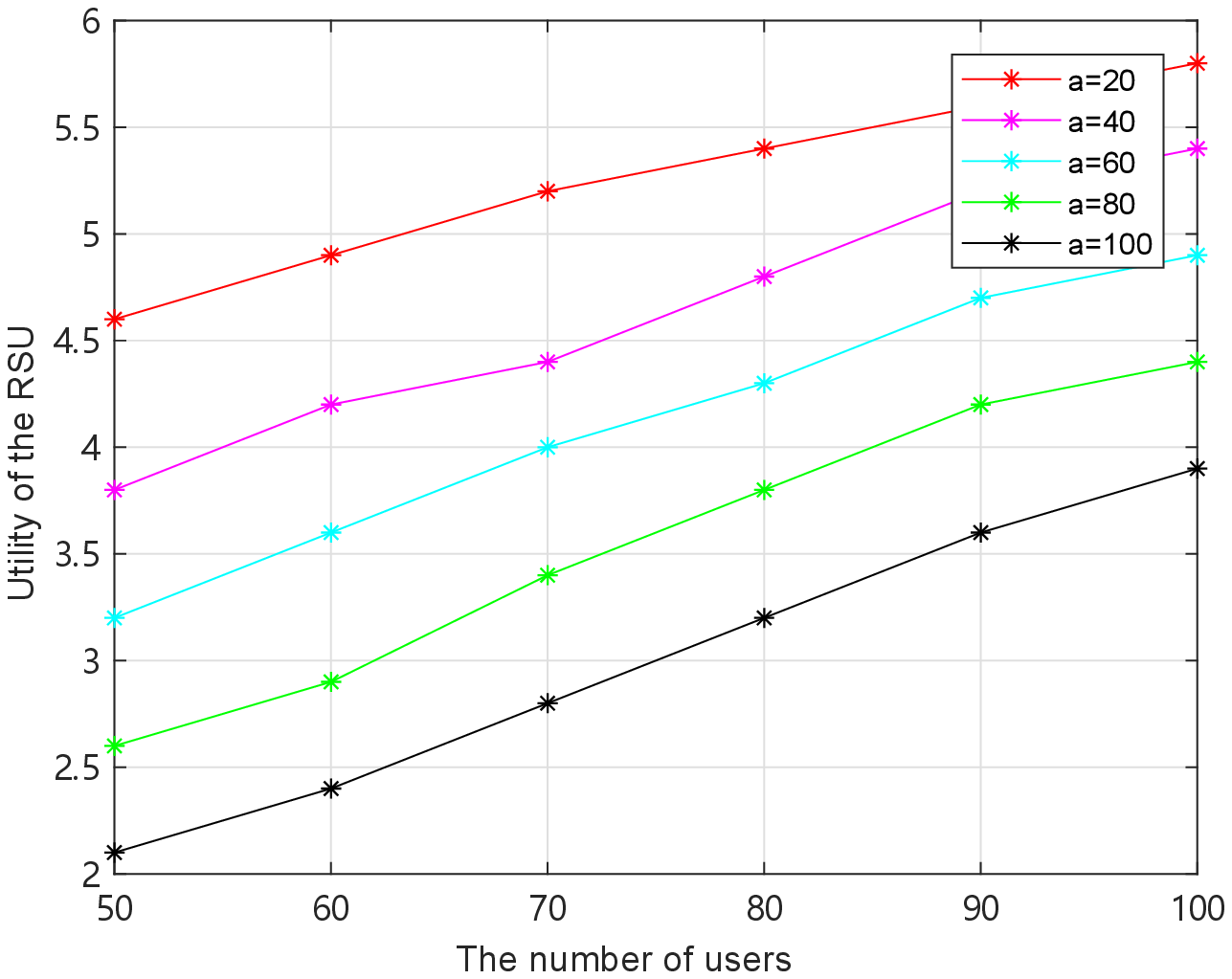}}
    \caption{Utility of the RSU versus the number of users}
    \label{fig:6}
\end{figure}

\begin{figure}[!t]
    \centerline{\includegraphics[scale=0.55]{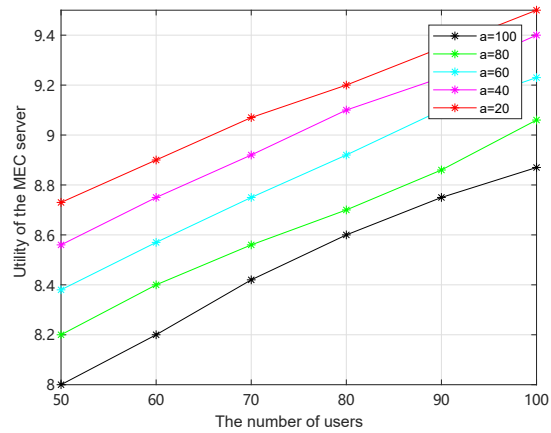}}
    \caption{Utility of the MEC server versus the number of users}
    \label{fig:7}
\end{figure}

Fig. \ref{fig:6} and Fig. \ref{fig:7} reveal the relationship between the
utility of RSU or MEC server and the number of users.  Fixed the parameter $a$,
it can be seen that with the number of users rising, the utility of RSU and MEC
server also rises.  However, when the number of users is fixed, with $a$
increasing, the utility of RSU and MEC server decreases.

\begin{figure}[!t]
    \centerline{\includegraphics[scale=0.55]{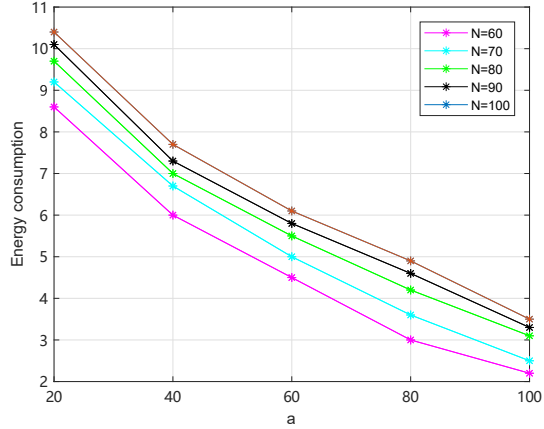}}
    \caption{Energy consumption of MEC server versus energy cost coefficient $a$}
    \label{fig:8}
\end{figure}

Fig. \ref{fig:8} shows that, with $a$ increasing, the energy consumption of MEC
server decreases.  This means that the MEC server tends to use external
resources.  Fixed $a$, with the number of users rising, the energy consumption
of MEC server also rises.  

\begin{figure}[!t]
    \centerline{\includegraphics[scale=0.55]{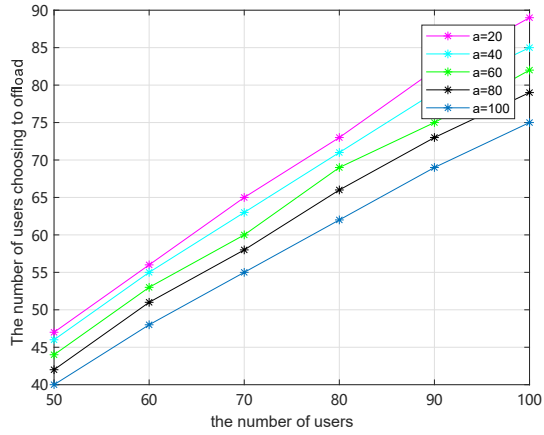}}
    \caption{The number of users choosing to offload versus the total number of
    users}
    \label{fig:9}
\end{figure}

Fig. \ref{fig:9} shows the relationship between the number of users choosing to
offload and the energy cost coefficient $a$.  We can see that, fixed the number
of users, when $a$ increases, since the energy consumption of MEC server
decreases, users must pay more for computing resources and tend not to offload.

\subsection{Comparative Study}

\begin{table*}\label{tab:2}
    \centering
    \caption{Performance comparison between our mechanism vs pure MEC offloading
    \cite{8917559}, cloud-MEC collaborative offloading \cite{8745530}, and
    autonomous MEC offloading \cite{8436044}}
    \begin{tabular}{l|l|l|l|l}
    \hline
    Compared Based On &Our Mechanism &Pure MEC Offloading &Cloud-MEC
    Collaborative Offloading &Autonomous MEC Offloading\\
    \hline
    \hline
    Delay              & low           & medium                & high   & low \\
    Energy Consumption & low           & high                  & medium & low \\
    Availability       & almost always & within server's range & always & always but not reliable \\
    Cost               & low           & medium                & high   & low \\
    \hline
    \end{tabular}
\end{table*}

\begin{figure}[!t]
    \centerline{\includegraphics[scale=0.55]{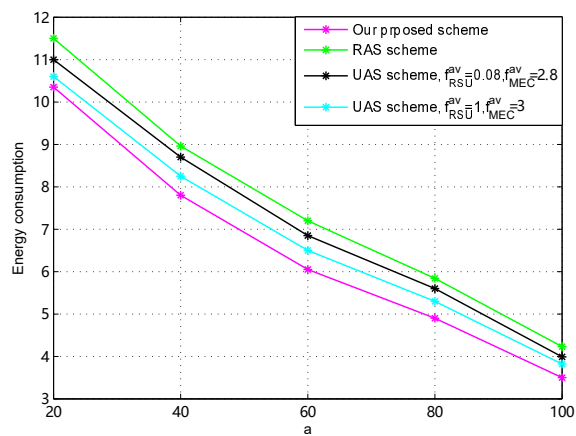}}
    \caption{Energy consumption of MEC server in different schemes}
    \label{fig:10}
\end{figure}

First, we compare our approach with the existing schemes for vehicular MEC
offloading.  That is, the pure MEC offloading with dedicated MEC server in
\cite{8917559}, the cloud-MEC collaborative computation offloading in
\cite{8745530}, where an MEC server and a cloud server coexist, and the
autonomous MEC offloading using neighbouring vehicles' idle computing resources
in \cite{8436044}.  In table \uppercase\expandafter {\romannumeral2}, we present
the performance comparison between our proposed mechanism and the existing pure
MEC offloading, cloud-MEC collaborative offloading and autonomous MEC
offloading, which shows the advantages of our proposed mechanism in offloading
delay, energy consumption and cost.  

Then, we compare our proposed optimal computing resource allocation scheme with
random allocation scheme (RAS) and uniform allocation scheme (UAS).  Given $N =
100$ users, RAS randomly allocates computing resources of the MEC server or
vehicles, while in UAS, we have $f_{RSU}^{av} = \{0.08, 0.1\} \mathrm{GHz},
f_{MEC}^{av} = \{2.8, 3\} \mathrm{GHz}$ for each user.  As shown in Fig.
\ref{fig:10}, the energy consumption of MEC server in our scheme is lower than
that in RAS and UAS schemes.  

\section{Conclusion}

In this paper, we propose a vehicular fog-edge computing network for computation
offloading.  In order to portray this scene, we formulate it as a multi-stage
Stackelberg game and a contract-based incentive mechanism.  We solve this
complex three-stage optimization problem by the backward induction method.  Then
we obtain the optimal price and computing resource demand strategies in each
stage.  Numerical results demonstrate the effectiveness of our proposed
incentive mechanism and trading mechanism, reveal the trends of energy
consumption and offloading decisions on various parameters, and show the
performance comparison between our work and the other existing MEC offloading
framework in vehicular networks.  Our future work will include two aspects.
First, based on HVC framework, combined with specific vehicle-to-everything
applications (\emph{e.g.}, sensor data sharing and processing, use cases of
automatic driving or advanced driving), the proposed algorithm will be verified
with real objects.  Besides, taking the heterogeneity of mobile device users and
of vehicles, a finer-grained task assignment and resource scheduling algorithm
will be studied.  

\section*{Acknowledgment}

This work was supported by the National Key Research and Development Program of
China (Grant No.2018YFB1702300), and in part by the NSF of China (Grants No.
61731012, 62025305, 61933009 and 92167205).

\section*{Appendix}

\subsection{Proof of Proposition 1}

\begin{IEEEproof}
Obviously, the objective function of problem \eqref{eq:P5} is concave.
Rewrite \eqref{eq:P5} as
\begin{equation}\label{eq:P8}
    \begin{aligned}
        \min_{p,f_e,f_{RSU}}\quad &-\tau + \Delta p + k_e f_e^2 + c f_{RSU} \\
        \mathrm{s.t.}\ \mathbf{C6:} &p \ge 0 \\
        \mathbf{C7:} &0\le f_e \le f_e^{\mathrm{max}} \\
        \mathbf{C8:} &f_{RSU} \ge 0 \\
        \mathbf{\overline{C9}:} &\frac{\tau}{p} - \Delta = f_e + f_{RSU}
    \end{aligned}
\end{equation}
The Lagrangian function is defined as
\begin{equation}
    \begin{aligned}
        \mathscr{L} &= -\tau + \Delta p + k_e f_e^2 + cf_{RSU} \\
        &+\alpha \left(\frac{\tau}{p} - \Delta - f_e - f_{RSU}\right) \\
        &+\gamma (f_e - f_e^{\mathrm{max}}) - \zeta p - \xi f_e - \nu f_{RSU}.
    \end{aligned}
\end{equation}
According to the Karush-Kunh-Tucker (KKT) conditions, we have
\begin{equation}\label{eq:partial-p}
    \frac{\partial \mathscr{L}}{\partial p} = \Delta - \frac{\alpha \tau}{p^2} -
    \zeta = 0,
\end{equation}
\begin{equation}\label{eq:partial-f_e}
    \frac{\partial \mathscr{L}}{\partial f_e} = 2 k_e f_e - \alpha + \gamma -
    \xi = 0,
\end{equation}
\begin{equation}\label{eq:partial-f_RSU}
    \frac{\partial \mathscr{L}}{\partial f_{RSU}} = c - \alpha - \nu = 0,
\end{equation}
\begin{equation}\label{eq:C9}
    \frac{\tau}{p} - \Delta - f_e - f_{RSU} = 0,
\end{equation}
\begin{equation}\label{eq:C7right}
    \gamma (f_e - f_e^{\mathrm{max}}) = 0,
\end{equation}
\begin{equation}
    \zeta p = 0,\ \xi f_e = 0,\ \nu f_{RSU} = 0,
\end{equation}
\begin{equation}
    p, f_e, f_{RSU}, \alpha, \gamma, \zeta, \xi, \nu \ge 0.
\end{equation}

\begin{lemma}
    $\zeta = 0$
\end{lemma}
\begin{IEEEproof}
    Suppose $\zeta \neq 0$.  Since $\zeta p = 0$, it follows that $p = 0$.
    However, if $p = 0$, then \eqref{eq:C9} does not hold.  Thus, we have $\zeta
    = 0$.  
\end{IEEEproof}

\begin{lemma}
    $\xi = 0$
\end{lemma}
\begin{IEEEproof}
    Suppose $\xi \neq 0$.  Similarly, we have $f_e = 0$.  If $f_e = 0$, from
    \eqref{eq:C7right} we know $\gamma = 0$.  From \eqref{eq:partial-f_e}, we
    have $\alpha = - \xi$.  Substituting $\alpha < 0$ into $\frac{\partial
    \mathscr{L}}{\partial p}$, we can obtain that $\frac{\partial \mathscr{L}}
    {\partial p} > 0$, which contradicts \eqref{eq:partial-p}.  Thus, $\xi = 0$.
\end{IEEEproof}

\begin{lemma}
    $\nu = 0$
\end{lemma}
\begin{IEEEproof}
    Suppose $\nu \neq 0$.  Then we have $f_{RSU} = 0$.  It means that the MEC
    server does not buy any computing resource from RSU.  Therefore, the price
    of RSU's resources is $c = 0$.  When $c = 0$, from \eqref{eq:partial-f_RSU}
    we can obtain that $\alpha = - \nu < 0$, which creates contradiction like
    Lemma 8.  Thus, we have $\nu = 0$.
\end{IEEEproof}

According to Lemma 7, 8 and 9, we can obtain the optimal solution of
\eqref{eq:P8} by the KKT conditions easily.  This problem is based on the
assumption that all users choose to offload their tasks.  Therefore, the price
of MEC server's computing resources must follow this condition:
\begin{equation}
    p < \min \left\{\frac{\tau_1}{f_1^l\delta + f_1^{thresh}}, \frac{\tau_2}
    {f_2^l\delta + f_2^{thresh}}, \cdots, \frac{\tau_N}{f_N^l\delta +
    f_N^{thresh}}\right\},
\end{equation}
From \eqref{eq:sort-users}, we can reduce this condition to $p < \frac{\tau_N}
{f_N^l\delta + f_N^{thresh}}$.  Then, from \eqref{eq:optimal-p}, we have
$\sqrt{\frac{c \sum_{i=1}^{N} \tau_i} {\delta \sum_{i=1}^{N} f_i^l}} <
\frac{\tau_N}{f_N^l\delta + f_N^{thresh}}$.  Thus, it follows that $c < T_N$.
Then, the optimal solution of \eqref{eq:P5} can be solved.  
\end{IEEEproof}

\subsection{Proof of Proposition 2}

\begin{IEEEproof}
The sufficiency part has been proved in the proof of Proposition 1.  Now we
consider the necessity part.  Suppose $T_N \le c < T_{N-1}$, and the optimal
price of MEC server's computing resources is $p^\star$ given by
\eqref{eq:optimal-p2}.  Since $c \ge T_N$, we have $p \ge \frac{\tau_N}{f_N^l
\delta + f_N^{thresh}}$ and $f_N^\star = 0$.  Then \eqref{eq:P4} can be
rewritten as
\begin{equation}\label{eq:P9}
    \begin{aligned}
        \max_{p,f_e,f_{RSU}}\quad &\sum_{i = 1}^{N-1} (\tau_i - f_i^l \delta p)
        - k_e f_e^2 - c f_{RSU} \\
        \mathrm{s.t.}\ \mathbf{C6}&\mathbf{-C8} \\
        \mathbf{\overline{C9}}&\mathbf{:}\ \sum_{i=1}^{N-1} \left( \frac{\tau_i}
        {p} - f_i^l \delta\right) = f_e + f_{RSU}
    \end{aligned}
\end{equation}

This problem has the same structure as \eqref{eq:P5}.  Therefore, from the proof
of the sufficiency part, we can see that the optimal price of MEC server's
computing resources for this problem is
\begin{equation}
    p^\star = \sqrt{\frac{c\sum_{i=1}^{N-1} \tau_i}{\delta \sum_{i=1}^{N-1}
    f_i^l}}.
\end{equation}
We can also obtain the optimal quantity of resources purchased by the MEC server
from RSU.

Obviously, the optimal solution of \eqref{eq:P9} is different from that of
\eqref{eq:P5}.  This contradicts with our presumption.  Thus, $p^\star$ given by
\eqref{eq:optimal-p} is the optimal solution of \eqref{eq:P5} only if $c < T_N$.
The necessity part is thus proved.  
\end{IEEEproof}

\subsection{Proof of Lemma 3}

\begin{IEEEproof}
According to IR conditions, we have
\begin{equation}
    \theta_1 p_1 - k_v f_1^2 \ge 0.
\end{equation}
Due to IC conditions, we have
\begin{equation}
    \theta_m p_m - k_v f_m^2 \ge \theta_m p_1 - k_v f_1^2, \forall m \in
    \{2,3,\cdots,M\}.
\end{equation}
As we know, $\theta_1 \le \theta_2 \le \cdots \le \theta_M$, thus
\begin{equation}
    \theta_m p_m - k_v f_m^2 \ge \theta_1 p_1 - k_v f_1^2 \ge 0, \forall m \in
    \{2,3,\cdots,M\}.
\end{equation}
That is, if the IR condition of type-$1$ vehicles is satisfied, the other IR
conditions also hold.  
\end{IEEEproof}

\subsection{Proof of Lemma 4}

\begin{IEEEproof}
According to IC conditions, we have
\begin{equation}\label{eq:temp1}
    \theta_{m+1}p_{m+1}-k_v f_{m+1}^2 \ge \theta_{m+1}p_m - k_v f_m^2,
\end{equation}
\begin{equation}\label{eq:temp2}
    \theta_m p_m - k_v f_m^2 \ge \theta_m p_{m-1} - k_v f_{m-1}^2.
\end{equation}
According to Lemma 1 and 2, we have
\begin{equation}\label{eq:temp3}
    \theta_{m+1}(p_m - p_{m-1}) \ge \theta_m (p_m - p_{m-1}).
\end{equation}
Combining \eqref{eq:temp2} and \eqref{eq:temp3}, we get
\begin{equation}\label{eq:temp4}
    \theta_{m+1}p_m - k_vf_m^2 \ge \theta_{m+1} p_{m-1} - k_v f_{m-1}^2.
\end{equation}
Combining \eqref{eq:temp1} and \eqref{eq:temp4}, we get
\begin{equation}
    \theta_{m+1}p_{m+1} - k_v f_{m+1}^2 \ge \theta_{m+1}p_{m-1} - k_v f_{m-1}^2.
\end{equation}
Recursively like this, we can prove Lemma 4.  
\end{IEEEproof}

\subsection{Proof of Lemma 5}

\begin{IEEEproof}
Assume that an LDIC constraint is not active, \emph{i.e.}, for type-$m$
vehicles,
\begin{equation}\label{eq:proofOfLemma5}
    \theta_m p_m - k_v f_m^2 > \theta_m p_{m-1} - k_v f_{m-1}^2.
\end{equation}
At this time, the RSU can gradually reduce the payment $p_m$ until both sides of
\eqref{eq:proofOfLemma5} equal.  This measure does not violate the LDIC
constraints, but improves the utility of RSU.  Therefore, under the optimal
contract, all the LDIC constraints must be active.

Similarly, we can prove that the IR constraint of type-$1$ vehicles is active as
well.  
\end{IEEEproof}

\subsection{Proof of Lemma 6}

\begin{IEEEproof}
According to Lemma 1, we have
\begin{equation}
    \theta_m (p_m - p_{m-1}) > \theta_{m-1} (p_m - p_{m-1}).
\end{equation}
From Lemma 5, we get
\begin{equation}
    k_v (f_m^2 - f_{m-1}^2) = \theta_m (p_m - p_{m-1}).
\end{equation}
Then,
\begin{equation}
    \begin{aligned}
        &k_v (f_m^2 - f_{m-1}^2) > \theta_{m-1} (p_m - p_{m-1}) \\
        \Rightarrow\ &\theta_{m-1}p_{m-1} - k_v f_{m-1}^2 > \theta_{m-1} p_m -
        k_v f_m^2,
    \end{aligned}
\end{equation}
\emph{i.e.}, the LUIC constraints hold.  
\end{IEEEproof}

\subsection{Proof of Theorem 2}

\begin{IEEEproof}
The first-order conditions of Lagrangian function \eqref{eq:Lagrangian} are
\begin{equation}\label{eq:fom}
    \frac{\partial \mathscr{L}}{\partial p_m} = \mu_{m+1}\theta_{m+1} - \mu_m
    \theta_m + l_m \theta_m = 0,\ \forall m \in \{2,3,\cdots,M - 1\},
\end{equation}
\begin{equation}\label{eq:fo1}
    \frac{\partial \mathscr{L}}{\partial p_1} = l_1 \theta_1 + \gamma \theta_1 +
    \mu_2 \theta_2 = 0,
\end{equation}
\begin{equation}\label{eq:foM}
    \frac{\partial \mathscr{L}}{\partial p_M} = l_M \theta_M - \mu_M \theta_M =
    0.
\end{equation}
Combining \eqref{eq:fom} and \eqref{eq:foM}, we can obtain \eqref{eq:mu_m}.  From
\eqref{eq:fo1}, we know that $\gamma = - \l_1 - \mu_2 \frac{\theta_2}
{\theta_1}$.  
\begin{equation}\label{eq:fofm}
    \frac{\partial \mathscr{L}}{\partial f_m} = 2 k_v (\mu_m - \mu_{m+1})f_m -
    l_m \eta = 0,\ \forall m \in \{2,3,\cdots,M-1\},
\end{equation}
\begin{equation}\label{eq:fof1}
    \frac{\partial \mathscr{L}}{\partial f_1} = -2 k_v (\mu_2 + \gamma) f_1 -
    l_1 \eta = 0,
\end{equation}
\begin{equation}\label{eq:fofM}
    \frac{\partial \mathscr{L}}{\partial f_M} = 2 k_v \mu_M f_M - l_M \eta = 0.
\end{equation}
From \eqref{eq:fofm}, \eqref{eq:fof1} and \eqref{eq:fofM}, we can obtain
\eqref{eq:f_m}, \eqref{eq:f_1} and \eqref{eq:f_M}.  

According to the reduced IR constraint $\mathbf{\overline{C4}}$, the solution of
$p_1$ can be written as \eqref{eq:p_1}.  Then, given the solution $f_i^\star,\
\forall i \in \{1,2,\cdots,M\}$ and the reduced IC constraints
$\mathbf{\overline{C5}}$, we can get the optimal solution $p_i^\star, \forall i
\in \{1,2,\cdots,M\}$, which is shown in \eqref{eq:p_m}.  

However, the analytical solution of $c$ can not be obtained directly.  The
gradient method can be used to update the Lagrangian multipliers to approximate
to the optimal $c$, \emph{i.e.},
\begin{equation}
    c^{(l)} = \left[c^{(l-1)} - \frac{\partial \mathscr{L}}{\partial c^{(l-1)}}
    \right]_{T_k}^{T_{k-1}}.
\end{equation}
\end{IEEEproof}

\printbibliography

@ARTICLE{7486992,
  author={C. {Wang} and Y. {Li} and D. {Jin} and S. {Chen}},
  journal={IEEE Transactions on Intelligent Transportation Systems}, 
  title={On the Serviceability of Mobile Vehicular Cloudlets in a Large-Scale Urban Environment}, 
  year={2016},
  volume={17},
  number={10},
  pages={2960-2970},
  doi={10.1109/TITS.2016.2561293}
}

@ARTICLE{7422842,
  author={R. {Yu} and J. {Ding} and X. {Huang} and M. {Zhou} and S. {Gjessing} and Y. {Zhang}},
  journal={IEEE Transactions on Vehicular Technology}, 
  title={Optimal Resource Sharing in 5G-Enabled Vehicular Networks: A Matrix Game Approach}, 
  year={2016},
  volume={65},
  number={10},
  pages={7844-7856},
  doi={10.1109/TVT.2016.2536441}
}

@ARTICLE{7271119,
  author={C. {Shao} and S. {Leng} and Y. {Zhang} and A. {Vinel} and M. {Jonsson}},
  journal={IEEE Transactions on Vehicular Technology}, 
  title={Performance Analysis of Connectivity Probability and Connectivity-Aware MAC Protocol Design for Platoon-Based VANETs}, 
  year={2015},
  volume={64},
  number={12},
  pages={5596-5609},
  doi={10.1109/TVT.2015.2479942}
}

@ARTICLE{7553459,
  author={K. {Zhang} and Y. {Mao} and S. {Leng} and Q. {Zhao} and L. {Li} and X. {Peng} and L. {Pan} and S. {Maharjan} and Y. {Zhang}},
  journal={IEEE Access}, 
  title={Energy-Efficient Offloading for Mobile Edge Computing in 5G Heterogeneous Networks}, 
  year={2016},
  volume={4},
  number={},
  pages={5896-5907},
  doi={10.1109/ACCESS.2016.2597169}
}

@ARTICLE{7415983,
  author={X. {Hou} and Y. {Li} and M. {Chen} and D. {Wu} and D. {Jin} and S. {Chen}},
  journal={IEEE Transactions on Vehicular Technology}, 
  title={Vehicular Fog Computing: A Viewpoint of Vehicles as the Infrastructures}, 
  year={2016},
  volume={65},
  number={6},
  pages={3860-3873},
  doi={10.1109/TVT.2016.2532863}
}

@ARTICLE{6574874,
  author={W. {Zhang} and Y. {Wen} and K. {Guan} and D. {Kilper} and H. {Luo} and D. O. {Wu}},
  journal={IEEE Transactions on Wireless Communications}, 
  title={Energy-Optimal Mobile Cloud Computing under Stochastic Wireless Channel}, 
  year={2013},
  volume={12},
  number={9},
  pages={4569-4581},
  doi={10.1109/TWC.2013.072513.121842}
}

@ARTICLE{7130662,
  author={S. {Sardellitti} and G. {Scutari} and S. {Barbarossa}},
  journal={IEEE Transactions on Signal and Information Processing over Networks}, 
  title={Joint Optimization of Radio and Computational Resources for Multicell Mobile-Edge Computing}, 
  year={2015},
  volume={1},
  number={2},
  pages={89-103},
  doi={10.1109/TSIPN.2015.2448520}
}

@ARTICLE{7762913,
  author={C. {You} and K. {Huang} and H. {Chae} and B. {Kim}},
  journal={IEEE Transactions on Wireless Communications}, 
  title={Energy-Efficient Resource Allocation for Mobile-Edge Computation Offloading}, 
  year={2017},
  volume={16},
  number={3},
  pages={1397-1411},
  doi={10.1109/TWC.2016.2633522}
}

@ARTICLE{8496832,
  author={Y. {Dai} and D. {Xu} and S. {Maharjan} and Y. {Zhang}},
  journal={IEEE Transactions on Vehicular Technology}, 
  title={Joint Computation Offloading and User Association in Multi-Task Mobile Edge Computing}, 
  year={2018},
  volume={67},
  number={12},
  pages={12313-12325},
  doi={10.1109/TVT.2018.2876804}
}

@ARTICLE{7572018,
  author={Y. {Mao} and J. {Zhang} and K. B. {Letaief}},
  journal={IEEE Journal on Selected Areas in Communications}, 
  title={Dynamic Computation Offloading for Mobile-Edge Computing With Energy Harvesting Devices}, 
  year={2016},
  volume={34},
  number={12},
  pages={3590-3605},
  doi={10.1109/JSAC.2016.2611964}
}

@INPROCEEDINGS{7541539,
  author={J. {Liu} and Y. {Mao} and J. {Zhang} and K. B. {Letaief}},
  booktitle={2016 IEEE International Symposium on Information Theory (ISIT)}, 
  title={Delay-optimal computation task scheduling for mobile-edge computing systems}, 
  year={2016},
  volume={},
  number={},
  pages={1451-1455},
  doi={10.1109/ISIT.2016.7541539}
}

@ARTICLE{7870694,
  author={C. {Wang} and F. R. {Yu} and C. {Liang} and Q. {Chen} and L. {Tang}},
  journal={IEEE Transactions on Vehicular Technology}, 
  title={Joint Computation Offloading and Interference Management in Wireless Cellular Networks with Mobile Edge Computing}, 
  year={2017},
  volume={66},
  number={8},
  pages={7432-7445},
  doi={10.1109/TVT.2017.2672701}
}

@ARTICLE{8533343,
  author={T. X. {Tran} and D. {Pompili}},
  journal={IEEE Transactions on Vehicular Technology}, 
  title={Joint Task Offloading and Resource Allocation for Multi-Server Mobile-Edge Computing Networks}, 
  year={2019},
  volume={68},
  number={1},
  pages={856-868},
  doi={10.1109/TVT.2018.2881191}
}

@ARTICLE{8946743,
  author={T. {Koketsu Rodrigues} and K. {Suto} and N. {Kato}},
  journal={IEEE Transactions on Emerging Topics in Computing}, 
  title={Edge Cloud Server Deployment with Transmission Power Control through Machine Learning for 6G Internet of Things}, 
  year={2019},
  volume={},
  number={},
  pages={1-1},
  doi={10.1109/TETC.2019.2963091}
}

@ARTICLE{8847416,
  author={T. K. {Rodrigues} and K. {Suto} and H. {Nishiyama} and J. {Liu} and N. {Kato}},
  journal={IEEE Communications Surveys   Tutorials}, 
  title={Machine Learning Meets Computation and Communication Control in Evolving Edge and Cloud: Challenges and Future Perspective}, 
  year={2020},
  volume={22},
  number={1},
  pages={38-67},
  doi={10.1109/COMST.2019.2943405}
}

@ARTICLE{7981532,
  author={J. {Liu} and J. {Wan} and B. {Zeng} and Q. {Wang} and H. {Song} and M. {Qiu}},
  journal={IEEE Communications Magazine}, 
  title={A Scalable and Quick-Response Software Defined Vehicular Network Assisted by Mobile Edge Computing}, 
  year={2017},
  volume={55},
  number={7},
  pages={94-100},
  doi={10.1109/MCOM.2017.1601150}
}

@misc{liu2019vehicular,
  title={Vehicular Edge Computing and Networking: A Survey}, 
  author={Lei Liu and Chen Chen and Qingqi Pei and Sabita Maharjan and Yan Zhang},
  year={2019},
  eprint={1908.06849},
  archivePrefix={arXiv},
  primaryClass={eess.SP}
}

@ARTICLE{8917559,
  author={V. {Huy Hoang} and T. M. {Ho} and L. B. {Le}},
  journal={IEEE Communications Letters}, 
  title={Mobility-Aware Computation Offloading in MEC-Based Vehicular Wireless Networks}, 
  year={2020},
  volume={24},
  number={2},
  pages={466-469},
  doi={10.1109/LCOMM.2019.2956514}
}

@ARTICLE{8745530,
  author={J. {Zhao} and Q. {Li} and Y. {Gong} and K. {Zhang}},
  journal={IEEE Transactions on Vehicular Technology}, 
  title={Computation Offloading and Resource Allocation For Cloud Assisted Mobile Edge Computing in Vehicular Networks}, 
  year={2019},
  volume={68},
  number={8},
  pages={7944-7956},
  doi={10.1109/TVT.2019.2917890}
}

@ARTICLE{9091251,
  author={H. {Ke} and J. {Wang} and L. {Deng} and Y. {Ge} and H. {Wang}},
  journal={IEEE Transactions on Vehicular Technology}, 
  title={Deep Reinforcement Learning-Based Adaptive Computation Offloading for MEC in Heterogeneous Vehicular Networks}, 
  year={2020},
  volume={69},
  number={7},
  pages={7916-7929},
  doi={10.1109/TVT.2020.2993849}
}

@ARTICLE{8936356,
  author={Z. {Xiao} and X. {Dai} and H. {Jiang} and D. {Wang} and H. {Chen} and L. {Yang} and F. {Zeng}},
  journal={IEEE Internet of Things Journal}, 
  title={Vehicular Task Offloading via Heat-Aware MEC Cooperation Using Game-Theoretic Method}, 
  year={2020},
  volume={7},
  number={3},
  pages={2038-2052},
  doi={10.1109/JIOT.2019.2960631}
}

@ARTICLE{8985335,
  author={Y. {Wang} and P. {Lang} and D. {Tian} and J. {Zhou} and X. {Duan} and Y. {Cao} and D. {Zhao}},
  journal={IEEE Internet of Things Journal}, 
  title={A Game-Based Computation Offloading Method in Vehicular Multiaccess Edge Computing Networks}, 
  year={2020},
  volume={7},
  number={6},
  pages={4987-4996},
  doi={10.1109/JIOT.2020.2972061}
}

@ARTICLE{8753694,
  author={B. {Gu} and Z. {Zhou}},
  journal={IEEE Vehicular Technology Magazine}, 
  title={Task Offloading in Vehicular Mobile Edge Computing: A Matching-Theoretic Framework}, 
  year={2019},
  volume={14},
  number={3},
  pages={100-106},
  doi={10.1109/MVT.2019.2902637}
}

@ARTICLE{8436044,
  author={G. {Qiao} and S. {Leng} and K. {Zhang} and Y. {He}},
  journal={IEEE Communications Magazine}, 
  title={Collaborative Task Offloading in Vehicular Edge Multi-Access Networks}, 
  year={2018},
  volume={56},
  number={8},
  pages={48-54},
  doi={10.1109/MCOM.2018.1701130}
}

@ARTICLE{8581401,
  author={J. {Feng} and Z. {Liu} and C. {Wu} and Y. {Ji}},
  journal={IEEE Vehicular Technology Magazine}, 
  title={Mobile Edge Computing for the Internet of Vehicles: Offloading Framework and Job Scheduling}, 
  year={2019},
  volume={14},
  number={1},
  pages={28-36},
  doi={10.1109/MVT.2018.2879647}
}

@ARTICLE{8926369,
  author={F. {Tang} and Y. {Kawamoto} and N. {Kato} and J. {Liu}},
  journal={Proceedings of the IEEE}, 
  title={Future Intelligent and Secure Vehicular Network Toward 6G: Machine-Learning Approaches}, 
  year={2020},
  volume={108},
  number={2},
  pages={292-307},
  doi={10.1109/JPROC.2019.2954595}
}

@ARTICLE{6786409,
  author={F. {Malandrino} and C. {Casetti} and C. {Chiasserini} and C. {Sommer} and F. {Dressler}},
  journal={IEEE Transactions on Vehicular Technology}, 
  title={The Role of Parked Cars in Content Downloading for Vehicular Networks}, 
  year={2014},
  volume={63},
  number={9},
  pages={4606-4617},
  doi={10.1109/TVT.2014.2316645}
}

@ARTICLE{7747510,
  author={Z. {Su} and Q. {Xu} and Y. {Hui} and M. {Wen} and S. {Guo}},
  journal={IEEE Transactions on Vehicular Technology}, 
  title={A Game Theoretic Approach to Parked Vehicle Assisted Content Delivery in Vehicular Ad Hoc Networks}, 
  year={2017},
  volume={66},
  number={7},
  pages={6461-6474},
  doi={10.1109/TVT.2016.2630300}
}

@ARTICLE{8383688,
  author={G. {Sun} and M. {Yu} and D. {Liao} and V. {Chang}},
  journal={IEEE Transactions on Intelligent Transportation Systems}, 
  title={Analytical Exploration of Energy Savings for Parked Vehicles to Enhance VANET Connectivity}, 
  year={2019},
  volume={20},
  number={5},
  pages={1749-1761},
  doi={10.1109/TITS.2018.2834569}
}

@incollection{grover2018vehicular,
  title={Vehicular fog computing paradigm: Scenarios and applications},
  author={Grover, Jyoti},
  booktitle={Vehicular Cloud Computing for Traffic Management and Systems},
  pages={200--215},
  year={2018},
  publisher={IGI Global}
}

@ARTICLE{8318667,
  author={X. {Wang} and Z. {Ning} and L. {Wang}},
  journal={IEEE Transactions on Industrial Informatics}, 
  title={Offloading in Internet of Vehicles: A Fog-Enabled Real-Time Traffic Management System}, 
  year={2018},
  volume={14},
  number={10},
  pages={4568-4578},
  doi={10.1109/TII.2018.2816590}
}

@ARTICLE{8463481,
  author={C. {Li} and S. {Wang} and X. {Huang} and X. {Li} and R. {Yu} and F. {Zhao}},
  journal={IEEE Internet of Things Journal}, 
  title={Parked Vehicular Computing for Energy-Efficient Internet of Vehicles: A Contract Theoretic Approach}, 
  year={2019},
  volume={6},
  number={4},
  pages={6079-6088},
  doi={10.1109/JIOT.2018.2869892}
}

@ARTICLE{9032323,
  author={H. {Sami} and A. {Mourad} and W. {El-Hajj}},
  journal={IEEE/ACM Transactions on Networking}, 
  title={Vehicular-OBUs-As-On-Demand-Fogs: Resource and Context Aware Deployment of Containerized Micro-Services}, 
  year={2020},
  volume={28},
  number={2},
  pages={778-790},
  doi={10.1109/TNET.2020.2973800}
}

@INPROCEEDINGS{7997316,
  author={Z. {Hou} and H. {Chen} and Y. {Li} and Z. {Han} and B. {Vucetic}},
  booktitle={2017 IEEE International Conference on Communications (ICC)}, 
  title={A contract-based incentive mechanism for energy harvesting-based Internet of Things}, 
  year={2017},
  volume={},
  number={},
  pages={1-6},
  doi={10.1109/ICC.2017.7997316}
}

@ARTICLE{1665007,
  author={B. {Johansson} and P. {Soldati} and M. {Johansson}},
  journal={IEEE Journal on Selected Areas in Communications}, 
  title={Mathematical decomposition techniques for distributed cross-layer optimization of data networks}, 
  year={2006},
  volume={24},
  number={8},
  pages={1535-1547},
  doi={10.1109/JSAC.2006.879364}
}

@ARTICLE{9277911,
  author={Shi, Jinming and Du, Jun and Wang, Jingjing and Wang, Jian and Yuan, Jian},
  journal={IEEE Transactions on Vehicular Technology}, 
  title={Priority-Aware Task Offloading in Vehicular Fog Computing Based on Deep Reinforcement Learning}, 
  year={2020},
  volume={69},
  number={12},
  pages={16067-16081},
  doi={10.1109/TVT.2020.3041929},
  ISSN={1939-9359},
  month={Dec},
}

@ARTICLE{9043503,
  author={Hou, Xiangwang and Ren, Zhiyuan and Wang, Jingjing and Cheng, Wenchi and Ren, Yong and Chen, Kwang-Cheng and Zhang, Hailin},
  journal={IEEE Internet of Things Journal}, 
  title={Reliable Computation Offloading for Edge-Computing-Enabled Software-Defined IoV}, 
  year={2020},
  volume={7},
  number={8},
  pages={7097-7111},
  doi={10.1109/JIOT.2020.2982292}
}

@article{bertsekas1997nonlinear,
  title={Nonlinear programming},
  author={Bertsekas, Dimitri P},
  journal={Journal of the Operational Research Society},
  volume={48},
  number={3},
  pages={334--334},
  year={1997},
  publisher={Taylor \& Francis}
}

\end{document}